\documentstyle[aps,prl,amsmath,epsfig]{revtex}
\tighten
\begin{document}
\draft
\twocolumn[\hsize\textwidth\columnwidth\hsize\csname@twocolumnfalse\endcsname
\title{Probing the Dark Energy with Quasar Clustering}
\author{M. O. Calv\~{a}o, J. R. T. de Mello Neto,
and I. Waga}
\address{{Universidade Federal do Rio de Janeiro, \\
Instituto de F\'\i sica, \\
CEP 21945-970 Rio de Janeiro, RJ, Brazil}}
\date{\today}

\maketitle

\begin{abstract} We show, through Monte Carlo simulations, that the
Alcock-Paczy\'nski test, as applied to quasar clustering, is a
powerful tool to probe the cosmological density and equation of
state  parameters, $\Omega_{m0}$, $\Omega_{x0}$ and $w$. By taking
into account the effect of peculiar velocities upon the
correlation function we obtain, for the Two-Degree Field QSO
Redshift Survey (2QZ), the predicted confidence contours for the
cosmological constant ($w=-1$) and spatially flat
($\Omega_{m0}+\Omega_{x0}=1$) cases. It turns out that, for
$w=-1$, the test is especially sensitive to the difference
$\Omega_{m0}-\Omega_{\Lambda 0}$, thus being ideal to combine with
CMB results. We also find out that, for the flat case, it is
competitive with future supernova and galaxy number count tests,
besides being complementary to them.

\end{abstract}
 ]

\paragraph*{Introduction.} Recent investigations of type Ia
supernovae (SNe Ia) suggest that the expansion of the Universe is
accelerating, driven by some kind of negative-pressure dark energy
\cite{Riess99,Perlmutter99}. Independent evidence for the SNe Ia
results is provided by observations of cosmic microwave
back\-ground (CMB) anisotropies in combination with constraints on
the matter density parameter ($\Omega _{m0}$)
\cite{deBernardis00}. The exact nature, however, of this dark
energy is not well understood at present. Vacuum energy or a
cosmological constant ($\Lambda$) is the simplest explanation, but
attractive alternatives like a dynamical scalar field
(quintessence) \cite{Ratra88} have also been explored in the
literature. An important task nowadays in cosmology is thus to
find new methods that could directly quantify the amount of dark
energy present in the Universe as well as determine its equation
of state and time dependence. New methods may constrain different
regions of the parameter space and are usually subject to
different systematic errors, and they are therefore crucial to
cross-check (or complement) the SNe results.

The test we focus on here is the one suggested by Alcock and
Paczy\'nski (hereafter AP)\cite{Alcock79}, which has attracted a
lot of attention during the last years
\cite{Ryden95,Ballinger96,Matsubara96,Hui99,McDonald99}. In
particular, Popowski \emph{et al.} \cite{Popowski98} (hereafter
PWRO) extended a calculation by Phillips \cite{Phillipps94} of the
geometrical distortion of the QSO correlation function. They
suggested a simple Monte Carlo experiment to see what constraints
should be expected from the 2dF QSO Redshift Survey (2QZ) and the
Sloan Digital Sky Survey (SDSS). However, they did not estimate
the probability density in the parameter space and, as a
consequence, they could not notice that the test is in fact very
sensitive to the difference $\Omega_{m0}-\Omega_{\Lambda 0}$.
Further, they did not take into account the effect of peculiar
velocities, although they discussed its role arguing that it would
not overwhelm the geometric signal.

Our aim, in this Letter, is to show the feasibility of redshift
distortion (geometric + peculiar velocity) measurements to
constrain cosmological parameters, by extending the PWRO Monte
Carlo experiments and obtaining confidence regions in the ($\Omega
_{m0},\Omega _{\Lambda 0} $) and ($\Omega_{m0}, w$) planes. We
compare the expected constraints from the AP test, when applied to
the 2QZ survey, with those obtained by other methods. We include a
general dark energy component with equation of state
$P_{x}=w\,\rho_{x}$, with $w$ constant. Our analysis can be
generalized to dynamical scalar field cosmologies as well as to
any model with redshift dependent equation of state. Since most
quasars have redshift $z\lesssim 2$ we expect the test to be
useful in the determination of a possible redshift dependence of
the equation of state. We explicitly take into account the effect
of large-scale coherent peculiar velocities. Our calculations are
based on the measured 2QZ distribution function and we consider
best fit values for the amplitude ($r_0$) and exponent ($\gamma$)
of the correlation function as obtained by Croom {\it et al.}
\cite{Croom01}. In this work, we only consider the 2QZ survey
although the results can easily be generalized to SDSS.

{\it Alcock-Paczy\'nski test and quasar clustering.} We assume
that the geometry is described by the standard Robertson-Walker
metric. By a straightforward calculation for null geodesics, we
obtain the radial coordinate $R$ as a function of $z$:
\begin{equation}
a_0R = g(z) := \left\{\begin{array}{lr}
\sinh(\sqrt{\Omega_{k0}}I(z))
/(H_0\sqrt{\Omega_{k0}}),\; \Omega _{k0}>0, \\
I(z), \; \Omega_{k0}=0, \\
\sin(\sqrt{-\Omega_{k0}}I(z)) /(H_0\sqrt{-\Omega_{k0}}),\; \Omega
_{k0}<0,
\end{array}\right.
\end{equation}
where~$a_0$~is~the~present~scale~factor, $I(z):=
\int_{z'=0}^z[H_0/H(z')]dz'$, and the Hubble parameter is given by
$H(z)=H_{0}[\Omega _{m0}(1+z)^{3}+\Omega
_{x0}(1+z)^{3(1+w)}+\Omega _{k0}(1+z)^{2}]^{1/2}$.

Given two close point sources (e.g., quasars), with coordinates
$(z,\theta,\phi)$ and $(z+dz,\theta + d\theta, \phi + d\phi)$,
directly read off a catalogue, the real-space infinitesimal
comoving distance between them can be decomposed, in the distant
observer approximation we adopt, into contributions parallel and
perpendicular to the line of sight, $r_{\perp } :=g(z)d\alpha$,
$r_{||} :={dz}/{H(z)}$, such that $r^{2}=r_{||}^{2}+r_{\perp
}^{2}$. Here, $d\alpha$ is the small angle between the lines of
sight.

The gist of the AP test relies then on the fact that, if we
observe an intrinsically spherical system ($r_{||}=r_{\perp }$),
it will appear distorted, in redshift space, according to the
generic formula $ {r_{\perp }}/{r_{||}}=j(z){s_{\perp }}/{s_{||}},
$ where the anisotropy or distortion function $j(z)$ is defined by
$ j(z):={g(z)H(z)}/{z}.$ Here we have assumed a Euclidean geometry
for redshift space, that is, $s_{||}:=dz$, $s_{\perp }:=zd\alpha
$, and $s^{2}=s_{||}^{2}+s_{\perp }^{2}$.

Observations \cite{Croom01} suggest that, on scales $\sim 1 -
40h^{-1}$ Mpc, the real space correlation function for quasars is
reasonably well fitted by a power law, $ \xi (r)=\left(
{r}/{r_{0}}\right) ^{-\gamma }$, which leads, in redshift space,
to an anisotropic correlation function, $ \xi (s,\mu )=\left[
{s}/{s_{0}(z)}\right]^{-\gamma }\left[{\mu ^{2}+j^{2}(z)(1-\mu
^{2})}\right]^{-\gamma /2}$, where $\mu:=s_{||}/s$ and $
s_0(z):=r_0H(z). \label{s0} $

Peculiar velocities also induce distortions in the correlation
function which can be confused with those arising from the
cosmological geometric effect. It is important to take them into
account when comparing theory with observations. For the $(z,s)$
range we will consider, the influence of small-scale velocity
dispersions is likely to be weak \cite{Popowski98} and we neglect
it in our analysis. The most relevant effect to be considered is
due to large-scale coherent flows \cite{Kaiser87}. The linear
theory correlation function is given by
\cite{Hamilton92,Matsubara96}
\begin{eqnarray}
&&\xi _{L}(s,\mu )=\left[\left( 1+\frac{2\beta }{3}+\frac{\beta
^{2}}{5}\right) P_{0}(\overline{\mu })+\left( \frac{4\beta
}{3}+\frac{4\beta ^{2}}{7}\right)\right. \nonumber\\
& &\left. \times \frac{\gamma }{\gamma -3}P_{2}(\overline{\mu })
+\frac{8\beta ^{2}}{35}\frac{\gamma (\gamma +2)}{(5-\gamma
)(3-\gamma )}P_{4}(\overline{\mu })\right]\xi(r), \label{corr}
\end{eqnarray}
\noindent where the $P_{i}(\overline{\mu})$ are Legendre
polynomials, and $\overline{\mu }:=r_{||}/r$. As usual, $\beta(z)
:={f(z)}/{b(z)}$, $f(z):=-d\ln D/d\ln(1+z)$ is the linear growth
rate, and we adopt the following dependence for the bias
parameter, $ b(z)=1+\left[D(z=0)/{D(z)}\right] ^{m}\left( b_{0}-1
\right) . $ If $m=1$, we have Fry's number-conserving bias model
\cite{Fry96}. The case $m=0$ corresponds to a constant bias, and
we also use $m\simeq 1.7$ in our computations, which seems to be
more in accordance with an observed nonevolving clustering
\cite{Croom01}. For models where the dark energy is a cosmological
constant ($w=-1)$, we use the Heath solution for the growing mode
\cite{Heath77},
$D(z)=\frac{5}{2}\Omega _{m0}H(z)\int_{z}^{\infty }{(1+x)}/{%
H(x)^{3}}dx,$ and the following approximation for the growth rate
\cite{Lahav91}, $f(z)\simeq \Omega
_{m}^{4/7}(z)+\frac{1}{70}{\Omega _{\Lambda }(z)}\left[ 1+ {\Omega
_{m}(z)}/{2}\right].$ For flat models, Silveira and Waga
\cite{Silveira94} obtained an exact solution for the growing mode,
$D(z) = {_{2}F}_{1}\left[ -\frac{1}{3w},\frac{w-1}{2},1-
\frac{5}{6w};\frac{ 1-\Omega _{m0}}{\Omega _{m0}}(1+z)
^{3w}\right]/(1+z)$, where $_{2}F_{1}[a,b,c;x]$ is the
hypergeometric function. The growth rate can also be expressed in
terms of hypergeometric functions.

Following PWRO, we obtain, for the number of pairs expected in an
infinitely small bin within ($z, s,\mu$) and ($z+dz, s+ds,
\mu+d\mu$),
\begin{equation}
dN_{pairs}=-\frac{2\pi }{A}\!\left( \frac{180\,N_{Q}\,F(z)}{\pi
\,z}\right) ^{\!2}\![1+\xi_{L}(s,\mu )]\,s^{2}dz\,ds\,d\mu.
\label{dNpairs}
\end{equation}
\noindent Here $A$ is the area (in deg$^2$) of the survey, $N_Q$
is the total number of sources (quasars) in the survey, and $F(z)$
is the normalized distribution function.

Croom {\it et al.} \cite{Croom01}, assuming an Einstein-de Sitter
Universe ($\Omega_{m}=1,\Omega_{\Lambda }=0$), showed that the
quasar clustering amplitude $r_{0}$ appears to vary very little
over the entire redshift range of the 2QZ survey. They found
$r_{0}\simeq 4\,h^{-1}$ Mpc as their best fit, which remains
nearly constant in comoving coordinate. Therefore, we have $
s_{0}|_{EdS}(z):= H(z)|_{EdS}\,{r_{0}}=({4}/{3000})(1+z)^{3/2}.$
Following again PWRO, we use the fact that the total number of
correlated pairs, $N_{pairs}$, in the survey is model independent
to scale $s_0(z)$ to other cosmologies. It is straightforward to
show that $N_{pairs}\propto {s_{0}^{3}(z)}/{j^{2}(z)}$, and we use
$s_{0}(z) =({4}/{3000})(1+z)^{3/2} \left[ {j(z) }/{j|_{EdS}(z)
}\right] ^{2/3}$ as a fiducial redshift-space correlation length
for our simulations.

A particular model predicts a number of pairs $A_{i}$ in each bin
of a $(z,s,\mu )$ space. In a real (or simulated) situation the
data consist of $N_{i}$ pairs in $i$ bins. PWRO showed that for
typical surveys, such as SDSS and 2QZ, we are bound to be in the
``sparse regime'' or ``Poisson limit''. In this case we may treat
bins in $(z,s,\mu )$ space as independent and the probability of
detecting $N_{i}$ pairs in bin $i$, when $A_{i}$ are expected, is
$P(N_{i}|A_{i})={e^{-A_{i}}A_{i}^{N_{i}}}/{N_{i}!} $ Since the
bins are independent, the likelihood ${\cal L}$ of obtaining the
data given the model is simply the product, ${\cal L}
=\prod_{i}P(N_{i}|A_{i})$. For a typical 2QZ simulation, we
assumed: (i) the completed survey will comprise $N_Q=26000$
quasars in a total area $A=750$ deg$^2$; (ii) the Einstein-de
Sitter fiducial correlation function has $r_0=4h^{-1}$ Mpc and
$\gamma=1.6$; (iii) the bias model is determined by $b_0=1.5$ and
$m=1$. The linear binning we chose covered the ranges:
$0.4<z<2.6$, $2<s/s_0(z)<7$, and $0<\mu<1$, with 16 bins in $z$,
25 in $s/s_0(z)$, and 5 in $\mu$, making up a total of 2000 bins.
The maximization of the likelihood was carried out with {\sc
minuit} \cite{James94} and cross-checked with {\sc mathematica}.
The probability density function was built via a Gaussian kernel
density estimate, from typically 1000 runs for each ``true''
model.

\begin{figure}
\centering \hspace*{0.in} \epsfig{file=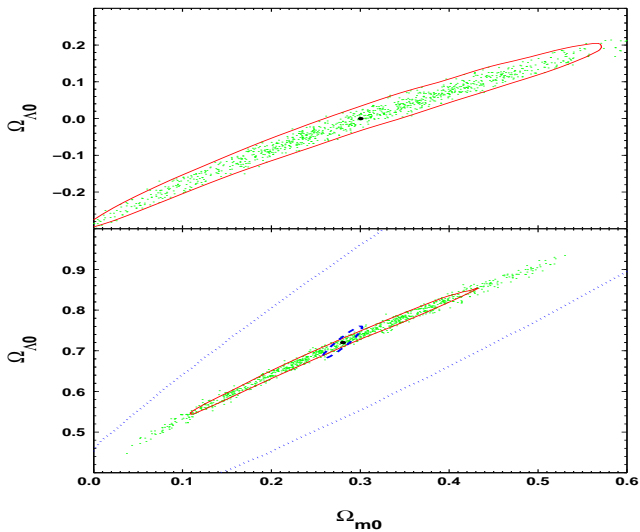,height= 7.0
cm,width= 8.5 cm} \vspace*{0.in} \caption{Simulated models at
fixed $w=-1$ and corresponding predicted AP confidence contours
(solid lines). In the top panel we show the predicted $2\sigma$
likelihood contour assuming a ``true'' model ($\Omega_{m0}=0.3,
\Omega_{\Lambda 0}=0)$. In the bottom panel the predicted
$1\sigma$ contour (dashed line) for one year of SNAP data
\protect\cite{Goliath01} is displayed, together with the predicted
$1\sigma$ AP contour. For both tests we consider
$\Omega_{m0}=0.28$ and $\Omega_{\Lambda 0}=0.72$; also displayed
is a $1\sigma$ confidence contour obtained by the Supernova
Cosmology Project (dotted lines;
\protect\cite{Perlmutter99}).}\label{fig:w-1}
\end{figure}

\paragraph*{Results and discussion.} In Figure \ref{fig:w-1}, we
show the predicted AP likelihood contours in the
($\Omega_{m0},\Omega_{\Lambda 0}$)-plane for the 2QZ survey (solid
lines), in the case $w=-1$, in a universe with arbitrary spatial
curvature. The scattered points represent maximum likelihood best
fit values for $\Omega_{m0}$ and $\Omega_{\Lambda 0}$. The assumed
``true'' values are ($\Omega_{m0}=0.3,\Omega_{\Lambda 0}=0$) and
($0.28,0.72$), for the top and bottom panels, respectively. In the
top panel the displayed curve corresponds to the predicted
$2\sigma$ likelihood contour. In the bottom panel the predicted
$1\sigma$ contour (dashed line) for one year of SNAP data
\cite{Goliath01} is displayed, together with the predicted
$1\sigma$ AP contour. For the SNAP contour, it is assumed that the
intercept $\cal{M}$ is exactly known. To have some ground of
comparison with current SNe Ia observations, in the same panel, we
also plot (dotted lines) the Supernova Cosmology Project
\cite{Perlmutter99} $1\sigma$ contour (fit C). As expected, in
both cases, the test recovers nicely the ``true'' values. We
stress out that the test is very sensitive to the difference
$\Omega_{m0}-\Omega_{\Lambda 0}$. From the bottom panel we note
that the sensitivity to this difference is comparable to that
expected from SNAP, of the order $\pm 0.01$. Comparatively,
however, the test has a larger uncertainty in the determination of
$\Omega_{m0}+\Omega_{\Lambda 0}$, of the order $\pm 0.21$. The
degeneracy in $\Omega_{m0}+\Omega_{\Lambda 0}$ may be broken if we
combine the estimated results for the AP test with, for instance,
those from CMB anisotropy measurements, whose contour lines are
orthogonal to those exhibited in the panels \cite{Hu99}.

\begin{figure}
\centering \hspace*{0.in} \epsfig{file=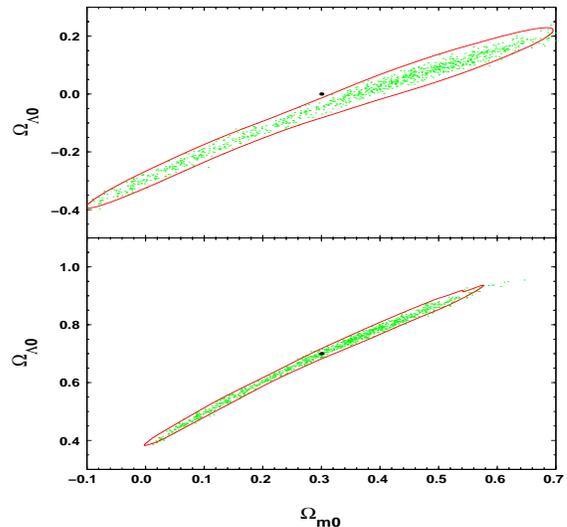,height= 7.0
cm,width= 7.5 cm} \vspace*{0.in} \caption{Simulated models at
fixed $w=-1$ and corresponding $2\sigma$ predicted AP confidence
contour; in both panels, the ``true'' model is indicated by a
solid dot. Top panel: The ``true'' model, $(0.3, 0)$, takes into
account the effect of peculiar velocities, but the simulated ones
do not. Notice that the ``true'' model does not fall into the
$2\sigma$ confidence region. Bottom panel: The ``true'' model,
$(0.3, 0.7)$, uses a redshift dependent bias function with
$b_0=1.45$ and $m=1.68$, whereas the simulated ones use a constant
bias equal to 2.46.}\label{fig:pvfavb}
\end{figure}

In order to estimate the consequences of neglecting the effect of
linear peculiar velocities, in the top panel of Figure
\ref{fig:pvfavb}, we included them in the calculation of the
$A_{i}$ values but neglected them in the computation of the
maximum likelihood; in this panel, we assume $\Omega _{m0}=0.3$
and $\Omega _{\Lambda 0}=0$ as ``true'' values. Notice that the
point with the ``true'' $\Omega _{m0}$ and $\Omega _{\Lambda 0}$
values is outside the $2\sigma $ contour. It is clear, therefore,
the necessity of taking this effect in consideration when
analyzing real data.

To illustrate that the AP test is in fact more sensitive to the
mean amplitude of the bias rather than to its exact redshift
dependence, we plot, in the bottom panel of Figure
\ref{fig:pvfavb}, the $2\sigma$ contour line, assuming as ``true''
values $\Omega_{m0}=0.3$ and $\Omega_{\Lambda 0}=0.7$. For this
panel, the ``true'' $A_i$ values were generated assuming
$b_0=1.45$ and $m=1.68$. However, for the simulations, we
considered a constant bias ($m=0$), such that
$b_{0,sim}:=\int_{z=z_{min}}^{z_{max}}F(z)b_{true}(z)dz=2.46$. We
remark that the contour is slightly enlarged and shifted in the
direction of the ``ellipsis'' major axis. However, the uncertainty
in $\Omega_{m0}-\Omega_{\Lambda 0}$ is practically unaltered,
confirming the strength of the test \cite{Yamamoto01}. We did the
same analysis assuming $\Omega_{m0}=1$ and $\Omega_{\Lambda 0}=0$
and obtained similar results.

\begin{figure}
\centering \hspace*{0.in} \epsfig{file=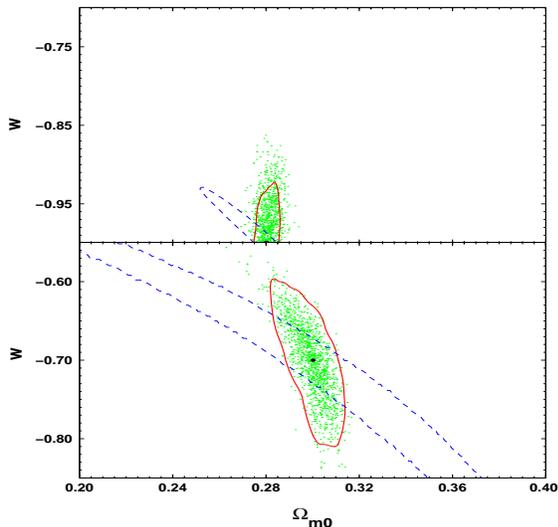,height= 7.0
cm,width= 7.5 cm} \vspace*{0.in} \caption{Simulated flat models
and corresponding predicted AP confidence contours (solid lines).
The top panel is from a ``true'' model ($\Omega_{m0}=0.28$,
$w=-1$), and displays the predicted confidence contours for the AP
test and the SNAP mission (dashed line; \protect\cite{Goliath01}),
both at $1\sigma$ level. The bottom panel is from a ``true'' model
($\Omega_{m0}=0.3$, $w=-0.7$), and displays the predicted
confidence contours for the AP test and the DEEP survey (dashed
line; \protect\cite{Newman00}), both at the $95\%$
level.}\label{fig:k0}
\end{figure}

In Figure \ref{fig:k0}, we show the predicted AP likelihood
contours in the $(\Omega_{m0},w)$-plane for the 2QZ survey (solid
lines) for flat models ($\Omega_{k0}=0$). The ``true'' values are
$(\Omega_{m0}=0.28,w=-1)$ and $(\Omega_{m0}=0.3,w=-0.7)$ for the
top and bottom panels, respectively. In the top panel, we show,
besides the AP contour, the predicted contour for one year of SNAP
data (dashed line; \cite{Goliath01}), both at $1\sigma$ level. For
the SNAP contour, the intercept $\cal{M}$ is assumed to be exactly
known. Notice that the contours are somewhat complementary and are
similar in strength. In the bottom panel, we compare the predicted
$95\%$ confidence contour of the AP test with the same confidence
contour for the number count test as expected from the DEEP
redshift survey (dashed line; \cite{Newman00}). Again the contours
are complementary, but the uncertainties on $\Omega_{m0}$ and $w$
for the AP test are quite smaller.

In summary, we have shown that the Alcock-Paczy\'nski test applied
to the 2dF quasar survey (2QZ) is a potent tool for measuring
cosmological parameters. We stress out that the test is especially
sensitive to $\Omega_{m0}-\Omega_{\Lambda 0}$. We have established
that the expected confidence contours are in general complementary
to those obtained by other methods and we again emphasize the
importance of combining them to constrain even more the parameter
space. We have also revealed that, for flat models, the estimated
constraints are similar in strength to those from SNAP with the
advantage that the 2QZ survey will soon be completed.

Of course our analysis can be improved in several aspects. For
instance, for the fiducial Einstein-de Sitter model, we have
assumed that $\gamma$ and $r_0$ do not depend on redshift. In
fact, observations \cite{Croom01} seem to support these
assumptions, but further investigations are necessary. Further, in
the simulations, for Figure \ref{fig:w-1} and Figure \ref{fig:k0},
we have assumed that the parameters $r_0$, $\gamma$, $b_0$ and $m$
are known exactly, that is, they are the same as the ``true''
input ones. Marginalization over these parameters is expected to
increase the size of the contours. However, preliminary results
where the errors in $r_0$ and $\gamma$ are taken into account
(supposed Gaussian), show that the confidence contours are not
appreciably altered. At present, the quasar clustering bias is not
completely well understood. Theoretical as well as observational
progress in its determination will certainly improve the real
capacity of the test. However, confirming previous investigations
\cite{Yamamoto01}, we have found that the test is, in fact, more
sensitive to the mean amplitude of the bias rather than to its
exact redshift dependence. A more extensive report of this work
and further investigations will be published elsewhere.

We would like to thank J. Silk for calling attention to the
potential of the AP test and T. Kodama for suggestions regarding
numerical issues. We also thank the Brazilian research agencies
CNPq, FAPERJ and FUJB.

\end{document}